\documentclass[reprint,notitlepage,onecolumn,showpacs,superscriptaddress,amsmath,amssymb]{revtex4-1}

\usepackage{amssymb}
\usepackage{amsmath}
\usepackage{enumerate}
\usepackage{graphicx}

\begin{document}

\title{Improving the efficiency of joint remote state preparation in noisy environment with weak measurement}

\author{Ming-Ming Wang}
\email{bluess1982@126.com}
\affiliation{School of Computer Science, Xi'an Polytechnic University, Xi'an 710048, China}
\affiliation{Jiangsu Engineering Center of Network Monitoring, Nanjing University of Information Science \& Technology, Nanjing 210044, China}
\author{Zhi-Guo Qu}
\affiliation{Jiangsu Engineering Center of Network Monitoring, Nanjing University of Information Science \& Technology, Nanjing 210044, China}

\date{\today}

\begin{abstract}
Quantum secure communication provides a new way for protecting the security of information. As an important component of quantum secure communication, remote state preparation (RSP) can securely transmit a quantum state from a sender to a remote receiver. The existence of quantum noise severely affects the security and reliability of quantum communication system. In this paper, we study the method for improving the efficiency of joint RSP (JRSP) subjected to noise with the help of weak measurement and its reversal measurement. Taking a GHZ based deterministic JRSP as an example, we utilize the technique of weak measurement and its reversal to suppress the effect of the amplitude-damping noise firstly. Our study shows that the fidelity of the output state can be improved in the amplitude-damping noise. We also study the effect of weak measurement and its reversal in other three types of noise usually encountered in real-world, namely, the bit-flip, phase-flip (phase-damping) and depolarizing noise. Our results show that the weak measurement has no effect for suppressing the bit-flip and phase-flip (phase-damping) noise, while has slight effect for suppressing the depolarizing noise. Our study is suitable for JRSP and RSP, and will be helpful for improving the efficiency of multiparticle entanglement based quantum secure communication in real implementation.
\end{abstract}


\maketitle

\section{Introduction}
Based on the basic principles of quantum mechanics, great progresses have been made in the field of communications and computation.
For one thing, quantum cryptograph \cite{BB84,HilleryBuzek-99,TerhalDivincenzo-998,CurtySantos-422,Wang-503}, especially
 quantum key distribution \cite{BB84}, has brought a new direction of cryptograph for achieving high-level security than their classical counterparts \cite{Shamir79,XiaWang-995,XiaWang-996,MaZhou-994,GuoWang-992,RenShen-993}.
For anther, quantum algorithms, like Shor's \cite{Shor97} and Grover’s \cite{Grover-225} algorithms, can solve a certain problem much faster than classical algorithms \cite{XiaWang-989,FuSun-1050,FuRen-1052}.

Entanglement is a crucial resource in quantum world and two amazing applications of entanglement are quantum teleportation \cite{BBC-93} and remote state preparation (RSP) \cite{Lo-54,Pati-50,BennettDivincenzo-53}, which can securely transmit a quantum state from a sender to a remote receiver by pre-shared entangled resource. If the preparer has known the information of the state, RSP can be performed with a simpler measurement and less classical communication costs. Since the first appearance of RSP, various types of RSP have been proposed, which include high-dimensional RSP \cite{Zeng-65}, oblivious RSP \cite{Leung-90}, faithful RSP \cite{YeZhang-69}, multicast RSP \cite{ZhouCheng-1071} and joint RSP (JRSP) \cite{Xia-40,Nguyen-41,Hou-48,Luo-283,Nguyen-283}, etc.
JRSP deals with the situation involving at least two preparers, where each preparer holds partial information and only if they work together can the state be prepared, which is suitable for protecting highly sensitive information.
However, a serious problem for most of these previous JRSP schemes \cite{Xia-40,Nguyen-41,Hou-48,Luo-283,Nguyen-283} is that the success probability is less than 1. To solve this problem, deterministic JRSP (DJRSP)\cite{XiaoLiu-386,NguyenCao-385,ChenXia-384,Wang-471} has been introduced by applying a three-step strategy. With some classical communication and local operations, the success probability of DJRSP can be increased to 1.

In a practical quantum communication system, the existence of quantum noise severely affects the security and reliability of the system \cite{WangWang-1001}. For an entanglement based quantum communication system, the pure entangled state shared among participants will be changed into a mixed one in the presence of noise. Some researches have studied the effect of noise in recent years.
Xiang et al. \cite{XiangLi-941} presented a RSP protocol for mixed state in depolarizing and dephasing channels. Chen et al. \cite{Ai-XiLi-942} investigated remote preparation of an entangled state through a mixed state channel in nonideal conditions. Guan et al. \cite{GuanChen-939} studied a JRSP of an arbitrary two-qubit state in the amplitude-damping and phase-damping noisy environment. Liang et al. \cite{LiangLiu-983,LiangLiu-1054} investigated a JRSP of a qubit state in different noises by solving Lindblad master equation.
Sharma et al. \cite{SharmaShukla-1057} investigated the effect of amplitude-damping and phase-damping noise on a bidirectional RSP protocol.
Li et al. \cite{LiLiu-1056} investigated a DJRSP of an arbitrary two-qubit state via four EPR pairs channel which are subjected to several Markovian noises and they analyzed the scheme by solving the master equation in Lindblad form.

Recently, the application of weak measurement \cite{AlbertVaidman-1084,UedaKoashi-1078} has been proposed as a practical method to protect the fidelity of quantum states subjected to decoherence through the amplitude-damping channel \cite{LeeJeong-1080,KimLee-1075,XiaAn-1076,PramanikMajumdar-1082,QiuTang-1083,RayChatterjee-1081}.
Some of these studies have shown that the technique of weak measurement with post selection and its reversal measurement can be employed to improve the fidelity of teleportation \cite{PramanikMajumdar-1082,QiuTang-1083} and quantum secret sharing \cite{RayChatterjee-1081}.
But these researches only investigated the amplitude-damping noise.
In this paper, we are going to study how to improve the efficiency of multiparticle entanglement based quantum secure communication in noisy environment using the weak measurement and its reversal measurement. Taking a GHZ based DJRSP scheme as an example, we utilize the technique of weak measurement and its reversal to suppress the effect of quantum noise. We investigate not only the amplitude-damping noise, but also the other three types of noise usually encountered in real-world, the bit-flip, phase-flip (phase-damping) and depolarizing noise.
The rest of this paper is organized as follows. In Sect. 2, the DJRSP scheme of an arbitrary one-qubit state is presented in noisy environment.
In Sect. 3, we investigate the DJRSP scheme in amplitude-damping noise with weak measurement and its reversal measurement.
In Sect. 4, we investigate effect of the weak measurement and its reversal measurement in the other three types of noise, respectively.
The paper is concluded in Sect. 5.

\section{DJRSP of an arbitrary one-qubit state in noisy environment}
In the following, a DJRSP scheme of an arbitrary one-qubit state based on GHZ state is given.  As is discussed in Ref. \cite{Wang-471}, this scheme is equivalent to the Bell state based scheme in Ref. \cite{NguyenCao-385}.

\subsection{DJRSP scheme of one-qubit based on GHZ state in the absence of noise}
In the DJRSP scheme, two preparers Alice and Bob will jointly prepare a qubit state for remote receiver Charlie. The prepared state has the form
\begin{equation} \label{EQ1}
{\left| \phi \right\rangle} = a_0 {\left| 0 \right\rangle}+ a_1 e^{\text{i} \theta} {\left| 1\right\rangle},
\end{equation}
where $a_0, a_1\in \mathcal{R}$ with $\sum^1_{j=0} a_j^2 = 1$,  $\theta \in [0, 2\pi]$.
The information of the prepared state is split as follows: Alice knows $ \{a_0, a_1\}$ and Bob knows $ \theta $.
A three-qubit GHZ state is shared among Alice, Bob and Charlie, which is
\begin{equation} \label{EQ2}
{\left| \text{GHZ}_3 \right\rangle} = \frac{1}{\sqrt{2}}({\left| 000 \right\rangle}+{\left|111 \right\rangle})_{\text{ABC}},
\end{equation}
where the subscripts denote the qubits of the state. Here, Alice holds qubit A, Bob holds B and Charlie holds C.

The DJRSP scheme runs as follows.

\textbf{Step 1:} Alice performs a projective measurement on qubit A in the basis  $\{ {\left| P_{m} \right\rangle}; m \in \{0,1\} \}$ with ${\left| P_{0}  \right\rangle} = a_{0} {\left| 0 \right\rangle} + a_{1}{\left| 1 \right\rangle}$, ${\left| P_{1}  \right\rangle} = a_{1} {\left| 0 \right\rangle} - a_{0}{\left| 1 \right\rangle}$.
Then, the quantum resource shared among three participants becomes
\begin{equation}\label{EQ3}
{\left| \text{GHZ}_3 \right\rangle}_{\text{ABC}}
 = \frac{1}{\sqrt{2}}  \sum^{1}_{m=0} {\left| P_{m} \right\rangle}_{\text{A}} {\left| Q_{m} \right\rangle}_{\text{BC}},
\end{equation}
with
${\left| Q_{0} \right\rangle}_{\text{BC}} =  a_0 {\left| 00 \right\rangle} + a_1 {\left| 11 \right\rangle}$,
${\left| Q_{1} \right\rangle}_{\text{BC}} =  a_1 {\left| 00 \right\rangle} - a_0 {\left| 11 \right\rangle}$.
After the measurement, Alice broadcasts $m$ to Bob and Charlie via classical channels.

\textbf{Step 2:}
Bob measures qubit B in the basis $\{ {\left| O_{m,n} \right\rangle}; ~m, n \in \{0,1\}\}$ that determined by both $\theta$ and $m$, where
\begin{equation} \label{EQ4}
\left(\begin{array}{l}
    {{\left| O_{0,0}\right\rangle} } \\
    {{\left| O_{0,1}\right\rangle} }  \end{array}\right)
=\frac{1}{\sqrt{2}}
\left(\begin{array}{cc}
 {1} &  { e^{- \text{i} \theta} } \\
 {1 } & {- e^{- \text{i} \theta} }
 \end{array}\right)
 \left(\begin{array}{l}
 {{\left| 0 \right\rangle} } \\
 {{\left| 1 \right\rangle} } \end{array}\right),
 ~~
 \left(\begin{array}{l}
    {{\left| O_{1,0} \right\rangle} } \\
    {{\left| O_{1,1} \right\rangle} }  \end{array}\right)
=\frac{1}{\sqrt{2}}
\left(\begin{array}{cc}
 { e^{- \text{i} \theta} }  & {1}  \\
 { -e^{- \text{i} \theta}} & {1}
 \end{array}\right)
 \left(\begin{array}{l}
 {{\left| 0 \right\rangle} } \\
 {{\left| 1 \right\rangle} } \end{array}\right).
\end{equation}

After Bob performs his measurement, ${\left| Q_{m} \right\rangle}$ can be rewritten as
\begin{equation}\label{EQ5}
{\left| Q_{m} \right\rangle}_{\text{BC}}
 = \frac{1}{\sqrt{2}}  \sum^{1}_{n=0} {\left| O_{m,n} \right\rangle}_{\text{B}} {R^{\dagger}_{m,n}} {\left| \phi \right\rangle}_{\text{C}},
\end{equation}
where $R_{m,n}$ denotes the recovery operator performed by Charlie.

\textbf{Step 3:} Bob announces $n$ publicly, then Charlie can perform $R_{m,n}$ on qubit C to get the prepared state ${\left| \phi \right\rangle}$, where $R_{0,0} = I $,  $R_{0,1} = \sigma_z$, $R_{1,0} = -\sigma_z\sigma_x$ and $R_{1,1} = -\sigma_x$

\subsection{The noise channels}
There are four types of noise usually encountered in real-world quantum communication protocols, namely the amplitude-damping, bit-flip, phase-flip (phase-damping) and depolarizing noise.

\subsubsection{The amplitude-damping noise}
The amplitude-damping noise describes the energy dissipation effects due to loss of energy from a quantum system and its Kraus operators are \cite{Xian-Ting-940}
\begin{equation} \label{EQ9}
        E_0=
        \left( \begin{array}{cc}
            1 & 0 \\
            0 & \sqrt{1-\lambda}
          \end{array}\right),
          ~~
          E_1=
        \left(\begin{array}{cc}
          0   &  \sqrt{\lambda} \\
          0  & 0
         \end{array}\right),
\end{equation}
where $0 \leq \lambda \leq 1$ indicates the noise parameter.

\subsubsection{The bit-flip noise}
The bit-flip noise changes the state of a qubit from ${\left| 0 \right\rangle}$ to ${\left| 1 \right\rangle}$ or from ${\left| 1 \right\rangle}$ to ${\left| 0 \right\rangle}$ with probability $\lambda$ and its Kraus operators are as follows \cite{Xian-Ting-940}
\begin{equation} \label{EQ6}
        E_0= \sqrt{1-\lambda}~I,
        ~~
        E_1= \sqrt{\lambda}~\sigma_x,
\end{equation}
where $I$ is identity matrix,  $\sigma_x$ is the Pauli matrix and $0 \leq \lambda \leq 1$ is the noise parameter.

\subsubsection{The phase-flip (phase-damping) noise}
The phase-flip noise changes the phase of the qubit ${\left| 1 \right\rangle}$  to $-{\left| 1 \right\rangle}$ with probability $\lambda$ and it can be described by Kraus operators as \cite{Xian-Ting-940}
\begin{equation} \label{EQ7}
        E_0= \sqrt{1-\lambda}~I,
        ~~
        E_1= \sqrt{\lambda}~\sigma_z,
\end{equation}
where $\sigma_z$ is the Pauli matrix and $0 \leq \lambda \leq 1$. Note that the phase-flip noise is equivalent to the phase-damping noise, which describes the loss of quantum information without energy dissipation.

\subsubsection{The depolarizing noise}

The depolarizing noise takes a qubit and replaces it with a completely mixed state $I/2$ with probability $\lambda$ and its Kraus operators are \cite{Xian-Ting-940}
\begin{equation} \label{EQ8}
\begin{split}
        E_0= \sqrt{1-\lambda}~I, ~~
        E_1= \sqrt{\frac{\lambda}{3}}~\sigma_x, ~~
        E_2= \sqrt{\frac{\lambda}{3}}~\sigma_z, ~~
        E_3= \sqrt{\frac{\lambda}{3}}~\sigma_y,
\end{split}
\end{equation}
where $\sigma_x, \sigma_z, \sigma_y$ are Pauli matrices and $0 \leq \lambda \leq 1$.

\subsection{Effect of noise on the DJRSP scheme}
\label{sec2-2}
An assumption of the above DJRSP scheme is that the GHZ state has been faithfully shared among three participants. But in real implementation, there must be a source who generates the entangled states and distributes qubits to relevant participants via quantum channels. And each distribution quantum channel will inevitably be affected by quantum noise.
Suppose Alice has a quantum source generator in her laboratory. She generates ${\left| \text{GHZ}_3 \right\rangle}_{\text{ABC}}$, keeps qubit A in her own and sends B to Bob and C to Charlie via noisy quantum channels, respectively.
In this case, the noisy effect on the entanglement after qubits transmissions can be represented as
\begin{equation}\label{EQ10}
\begin{split}
  \rho_\mathrm{noise}  = \sum_{j_1, j_2}
         { E_{j_1}^{\text{B}}  E_{j_2}^{\text{C}} }
         \left| \text{GHZ}_3 \right\rangle  \left\langle \text{GHZ}_3 \right|
         {E_{j_1}^{\text{B}}}^{\dag}  {E_{j_2}^{\text{C}}}^{\dag},
\end{split}
\end{equation}
where $E_{j_1}$ and $E_{j_2}$ are the noise operators, and superscripts B and C denote the qubits that are acted by noise operators.

In the presence of noise, Alice, Bob and Charlie share $\rho_\mathrm{noise}$ as the quantum resource to perform the DJRSP scheme. The process of the scheme is rewritten in the form of density operator as follows.

\textbf{Step 1:}  Alice measures qubit A by using operators $\{\mathcal{A}_m=  {\left| P_m \right\rangle}{\left\langle P_m  \right|};~m \in \{0,1\} \}$, and the system of (B, C) becomes
        \begin{equation}\label{EQ11}
        \begin{split}
        \rho^{\text{BC}}_{m}  =
        \frac{1}{p^{\text{A}}_{m} }
        \text{tr}_{\text{A}} \left( \mathcal{A}_m ~ \rho_\mathrm{noise}~  \mathcal{A}_m^{\dag} \right),
        \end{split}
        \end{equation}
where $ p^{\text{A}}_{m}$ is the probability that Alice gets $m$, which has the form
        \begin{equation}\label{EQ11-1}
        \begin{split}
        p^{\text{A}}_{m} = {\text{tr}(\mathcal{A}_m^{\dag}  \mathcal{A}_m ~\rho_\mathrm{noise})}.
        \end{split}
        \end{equation}

\textbf{Step 2:} Bob measures qubit B by using $\{ \mathcal{B}_{m,n} =  {\left|\right. O_{m,n} \left.\right\rangle}{\left\langle\right. O_{m,n}  \left.\right|}; ~m, n \in \{0,1\} \}$, and qubit C becomes
        \begin{equation}\label{EQ12}
        \begin{split}
        \rho^{\text{C}}_{m,n} & =
        \frac{1}{p^{\text{B}}_{m,n}}
       \text{tr}_{\text{B}}  \left( {\mathcal{B}_{m,n}} ~ \rho^{\text{BC}}_{m}  ~ {\mathcal{B}^{\dag}_{m,n}} \right),
        \end{split}
        \end{equation}
where $p^{\text{B}}_{m,n} $ is the probability that Bob gets $n$ on condition $m$, which has the form
        \begin{equation}\label{EQ12-1}
        \begin{split}
        p^{\text{B}}_{m,n}  = { \text{tr} \left( {\mathcal{B}^{\dag}_{m,n} } {\mathcal{B}_{m,n} } ~\rho^{\text{BC}}_{m}  \right)} .
        \end{split}
        \end{equation}

\textbf{Step 3:} Charlie recover the prepared state by performing $R_{m, n}$ on the qubit C, that is
       \begin{equation}\label{EQ13}
        \begin{split}
        \rho_\mathrm{out}^{m,n}  & = R_{m, n} ~\rho^{\text{C}}_{m,n} ~ {R^{\dag}_{m, n}}.
        \end{split}
        \end{equation}
It is clear that Charlie will get the output state as $\rho_\mathrm{out}^{m,n}$ depending on Alice's measurement result $m$ and Bob's $n$.

\subsection{The fidelity of the output state}
Since the effect of quantum noise, the output state $\rho_\mathrm{out}^{m,n}$ in Charlie's side might not be the desired prepared state $ {\left| \phi \right\rangle}$. Generally, fidelity can be used to describe the distance between these two states. For each $m$ and $n$, the fidelity of the output state is defined as
\begin{equation}\label{EQ15}
\begin{split}
F_{m,n} := | {\left\langle \phi \right|} \rho_\mathrm{out}^{m,n}   {\left| \phi \right\rangle}|.
\end{split}
\end{equation}

Note that each output state may occur with different probabilities. The average fidelity can be defined as
\begin{equation}\label{EQ16}
\begin{split}
\overline{F}  = \sum_{m,n}p^{\text{A}}_{m} p^{\text{B}}_{m,n}F_{m,n}.
\end{split}
\end{equation}

To eliminate parameters  $ a_0, a_1$ and $ \theta $ of the prepared state and compute the average fidelity over all possible prepared states, the state-independent average fidelity is defined as \cite{RigolinFortes-1033}
\begin{equation}\label{EQ17}
\begin{split}
\langle\overline{F}\rangle =  \frac{1}{2\pi}{\int_0^{2\pi} {\int_0^1  \overline{F} d a^2_1 d\theta} }.
\end{split}
\end{equation}

\subsection{The state-independent average fidelity in all types of noise}
To analyze noise effect of each type of noise, one need to calculate $\rho_{noise}$, put it into Eqs. (\ref{EQ11}) to (\ref{EQ13}), and get the state-independent average fidelity.
For the above four types of noise, one will get the state-independent average fidelity as follows.

In the amplitude-damping noise, the state-independent average fidelity is
\begin{equation}\label{EQ21}
\begin{split}
      \langle \overline{F_\mathrm{AD}} \rangle = 1-\frac{\lambda }{2}.
  \end{split}
  \end{equation}

In the bit-flip noise, one will get the fidelity as
\begin{equation}\label{EQ22}
\begin{split}
      \langle\overline{F_\mathrm{BF}}\rangle = \frac{2}{3} \lambda ^2-\lambda +1.
  \end{split}
  \end{equation}

In the phase-flip noise, the fidelity is
\begin{equation}\label{EQ23}
\begin{split}
     \langle\overline{F_\mathrm{PF}}\rangle   =  \frac{1}{3} \left(4 \lambda ^2-4 \lambda +3\right).
\end{split}
\end{equation}

In the depolarizing noise, the fidelity is
\begin{equation}\label{EQ24}
\begin{split}
     \langle \overline{F_\mathrm{DE}}\rangle =  \frac{16}{27} \lambda ^2 -\frac{10 }{9} \lambda+1.
  \end{split}
  \end{equation}

\section{DJRSP in the amplitude-damping noise with weak measurement and reversal measurement}

\subsection{DJRSP in the amplitude-damping noise with weak measurement and reversal measurement}

The weak measurement and its reversal measurement can be utilized to improve the fidelity of teleportation and RSP in noisy environment. In the DJRSP scheme, the weak measurements and its reversal can be applied in two places: one before and the other after the decoherence acts on the system.
The operator of the weak measurement can be described as,
\begin{equation} \label{EQ30}
        \mathcal{W}_0=
        \left( \begin{array}{cc}
            1 & 0 \\
            0 & \sqrt{1-s}
          \end{array}\right),
\end{equation}
where $s$ is the strength of  the weak measurement.
The weak measurement can be achieved by reducing the sensitivity of the detector, i.e., the detector clicks with probability $s$ if the input qubit is in $\left| 1 \right\rangle$, and never clicks if the input qubit is in $\left| 0 \right\rangle$.
The operator of the reverse weak measurement are given by
\begin{equation} \label{EQ31}
        \mathcal{V}_0=
        \left( \begin{array}{cc}
        \sqrt{1-r} & 0 \\
            0 & 1
          \end{array}\right),
\end{equation}
where $r$ is the strength of the reversal weak measurement.

The DJRSP of one-qubit state with weak measurement and its reversal is as follows.

\textbf{Step A1:}
Alice generates $\left| \text{GHZ}_3 \right\rangle$. She makes two weak measurements on qubits B and C before sending them through noisy channels. It is assumed that both weak measurements on qubits B and C have the same strength $s$. The system after this process can be described as
\begin{equation} \label{EQ32}
 \begin{split}
  \rho_{\mathcal{W}_0}  =
    \frac{  ( I   \otimes  \mathcal{W}_0  \otimes \mathcal{W}_0 )
                  ~\left| \text{GHZ}_3 \right\rangle  \left\langle \text{GHZ}_3 \right|~
                    ( I   \otimes  \mathcal{W}^{\dag}_0  \otimes \mathcal{W}^{\dag}_0)}
        {p_{\mathcal{W}_0} },
\end{split}
\end{equation}
where $p_{\mathcal{W}_0}$ is the probability of Alice gets $\mathcal{W}_0$ on qubits B and C, which has the form
\begin{equation}
\begin{split}
  p_{\mathcal{W}_0}
  & =
  \text{tr}\{( I   \otimes  \mathcal{W}^{\dag}_0  \otimes \mathcal{W}^{\dag}_0)  ( I   \otimes  \mathcal{W}_0  \otimes \mathcal{W}_0 )
         ~\left| \text{GHZ}_3 \right\rangle  \left\langle \text{GHZ}_3 \right| \}.
\end{split}
\end{equation}

\textbf{Step A2:}
Alice sends qubits B and C to Bob and Charlie through noisy channels, respectively. And the noise effect on the system is described as
\begin{equation}\label{EQ33}
\begin{split}
  \rho_\mathrm{noise}  = \sum_{j_1, j_2}
         { E_{j_1}^{\text{B}}  E_{j_2}^{\text{C}} }
         ~\rho_{\mathcal{W}_0}~
         {E_{j_1}^{\text{B}}}^{\dag}  {E_{j_2}^{\text{C}}}^{\dag}.
\end{split}
\end{equation}

\textbf{Step A3:}
After receiving qubits B and C from noisy channels, Bob and Charlie perform reversal quantum measurements on qubits B and C, respectively. Then, the system shared among Alice, Bob and Charlie becomes
\begin{equation} \label{EQ34}
 \begin{split}
  \rho_{\mathcal{V}_0}  =
  \frac{  ( I   \otimes  \mathcal{V}_0  \otimes \mathcal{V}_0 )
        ~\rho_\mathrm{noise}~
        ( I   \otimes  \mathcal{V}^{\dag}_0  \otimes \mathcal{V}^{\dag}_0)}
  {p_{\mathcal{V}_0}}.
\end{split}
\end{equation}
where $p_{\mathcal{V}_0}$ is the probability of Bob and Charlie get $\mathcal{V}_0 $, that is
\begin{equation} \label{EQ35}
 \begin{split}
  p_{\mathcal{V}_0}
  & = \text{tr} \{
      ( I   \otimes  \mathcal{V}^{\dag}_0  \otimes \mathcal{V}^{\dag}_0)
      ( I   \otimes  \mathcal{V}_0  \otimes \mathcal{V}_0 )
        ~\rho_\mathrm{noise}\}.
\end{split}
\end{equation}

\textbf{Step A4:}
In this case, Alice, Bob and Charlie use $\rho_{\mathcal{V}_0}$ as shared resource to perform the DJRSP scheme, which is shown in Eqs. (\ref{EQ11}) to (\ref{EQ13}), by replacing $\rho_{\text{noise}}$ with $\rho_{\mathcal{V}_0}$ in (\ref{EQ11}) and (\ref{EQ11-1}).

\textbf{Step 1-3:} Similar to Step 1-3 in Sect. \ref{sec2-2}.

\subsection{The state-independent average fidelity with weak measurement and reversal measurement}
In the amplitude-damping noise, one will get the state-independent average fidelity with weak measurement and reversal measurement as
\begin{equation}\label{EQ40}
\begin{split}
\langle\overline{F'_\mathrm{AD}}\rangle =
\frac{2 \left(r^2+r (s-3)+(s-3) s+3\right)+\lambda ^2 r (r+1) (s-1)^2-\lambda  (s-1) (r (3 s-1)+s-3)}{3 \left(r \left(\lambda ^2 r (s-1)^2+r-2 \lambda  (s-1)^2-2\right)+(s-2) s+2\right)}.
\end{split}
\end{equation}
which is related to the noise rate $\lambda$, the strength of the weak measurement $s$ and the strength of the reversal measurement $r$, but independent of the parameters of the prepared state.

The aim is to improve the state-independent average fidelity $\langle\overline{F_\mathrm{AD}}\rangle$ in Eq. \ref{EQ21}. To improve the state-independent average fidelity in the amplitude-damping environment, one need to choose the strengths of the weak measurement $s$ and the reversal measurement $r$ properly. The optimal strength can be obtained by maximizing Eq. (\ref{EQ40}) with respect to $r$. In this case, the optimal reversal measurement strength $r$ is
\begin{equation}
r_{\text{opt}}=\frac{\delta +(\delta -1) \lambda  (s-1)-\delta  s+\lambda ^2 (s-3) (s-1)^2-2}{\lambda  (s-1) (\lambda  (s-1) (\lambda  (s-1)-2)-1)-2},
\end{equation}
with $\delta =\sqrt{\lambda  (s-1) (5 \lambda  (s-1)+4)+4}$ on condition that $0 < s < 1$ and $ 0 \leq \lambda < 1$.

Thus, the optimal value of  $\langle\overline{F'_\mathrm{AD}}\rangle$ respect to $r_{\text{opt}}$ will be
\begin{equation}
\begin{split}
\langle\overline{F'_\mathrm{AD}}\rangle_{\text{opt}} =
\frac{1}{6} \left(4 -2 \lambda+2 \lambda  s +\sqrt{5 \lambda ^2-4 \lambda +5 \lambda ^2 s^2-10 \lambda ^2 s+4 \lambda  s+4}\right).
\end{split}
\end{equation}

\begin{figure}
\centering
\includegraphics [scale=0.50]{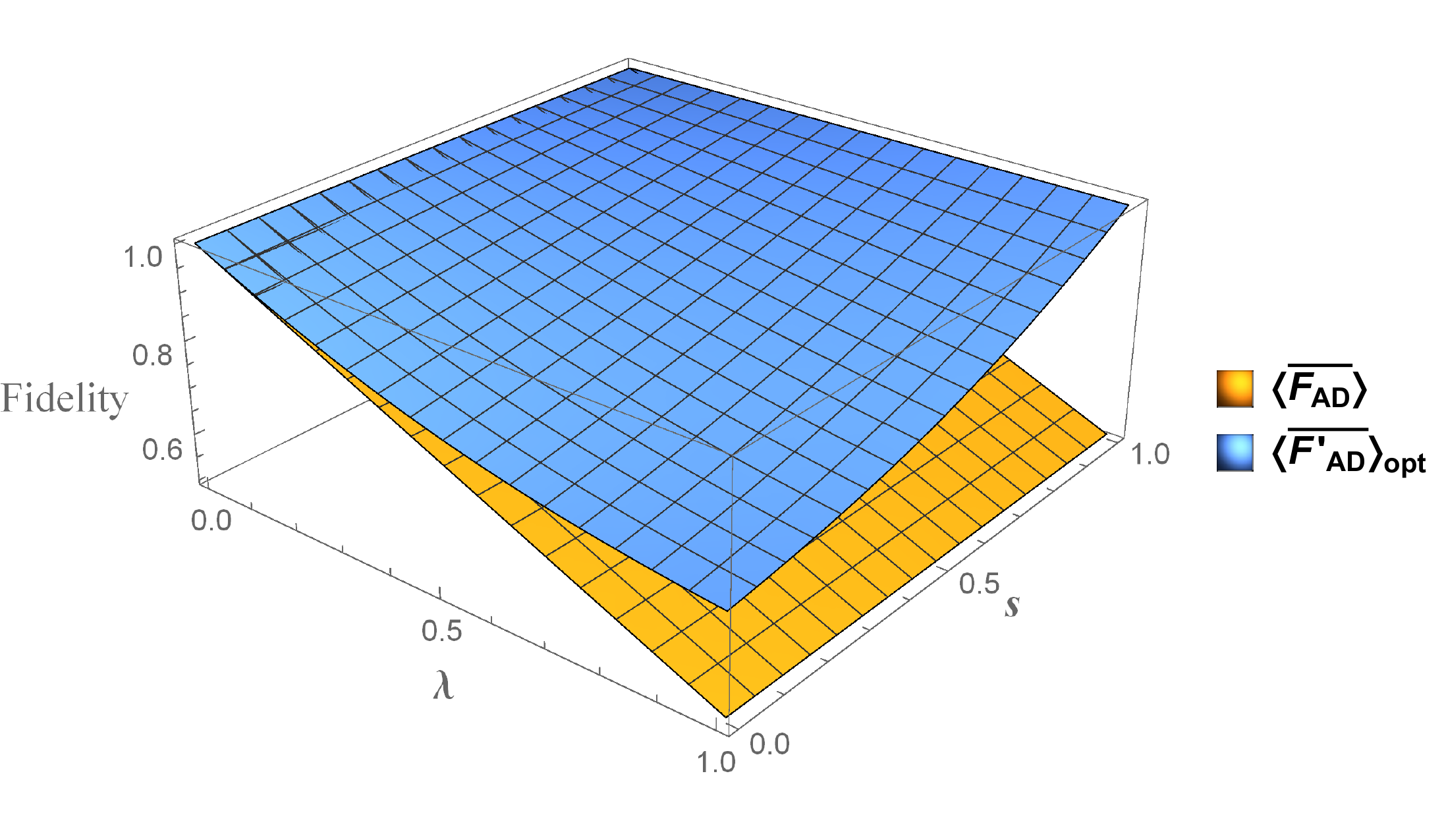}
  \caption{The state-independent average fidelities $\langle\overline{F_\mathrm{AD}}\rangle$ and $\langle\overline{F'_\mathrm{AD}}\rangle_{\text{opt}}$ with noise parameter $\lambda$ and weak measurement strength $s$ in the amplitude-damping noise, where $\langle\overline{F_\mathrm{AD}}\rangle$ is the fidelity without weak measurement and $\langle\overline{F'_\mathrm{AD}}\rangle_{\text{opt}}$ is the optimal  fidelity with weak and reversal measurements. }
    \label{fig:1}
\end{figure}

The state-independent average fidelity $\langle\overline{F_\mathrm{AD}}\rangle$ without weak measurement and the optimal  fidelity  $\langle\overline{F'_\mathrm{AD}}\rangle_{\text{opt}}$ with weak measurement and its reversal are plotted in Fig. \ref{fig:1}.
It can be seen from the figure that the optimal $\langle\overline{F'_\mathrm{AD}}\rangle_{\text{opt}}$ where weak measurement and its reversal performed is always larger than $\langle\overline{F_\mathrm{AD}}\rangle$ where no weak measurement performed, which means the state-independent average fidelity has been highly improved with weak measurement and its reversal.

\subsection{The success probability}
Although the state-independent average fidelity has been highly improved, there is a trade-off between the improvement of average fidelity and the total success probability. As the weak measurement and the reversal measurement are non-unitary operations, the scheme has less than 1 success probability. The success probability of the DJRSP scheme in the amplitude-damping noise with weak measurement and its reversal can be calculated as
\begin{equation}
  p_{\text{AD}} = p_{\mathcal{W}_0} *  p^{\text{AD}}_{\mathcal{V}_0}=\frac{\delta  (\lambda -1)^2 (s-1)^2}{\delta +\lambda  (s-1) (\lambda  (s-1) (\delta +2 \lambda  (s-1)+2)+2)},
\end{equation}
where $ p_{\mathcal{W}_0} $ is irrelevant to quantum noise since it has been performed before the qubits transmissions, while  $p^{\text{AD}}_{\mathcal{V}_0}$ is related to the noise rate of the amplitude-damping noise since it is the probability Bob and Charlie get $\mathcal{V}^{\text{AD}}_0 $ after qubits B and C have been transmitted through noise channels.

\begin{figure}
\centering
\includegraphics [scale=0.55]{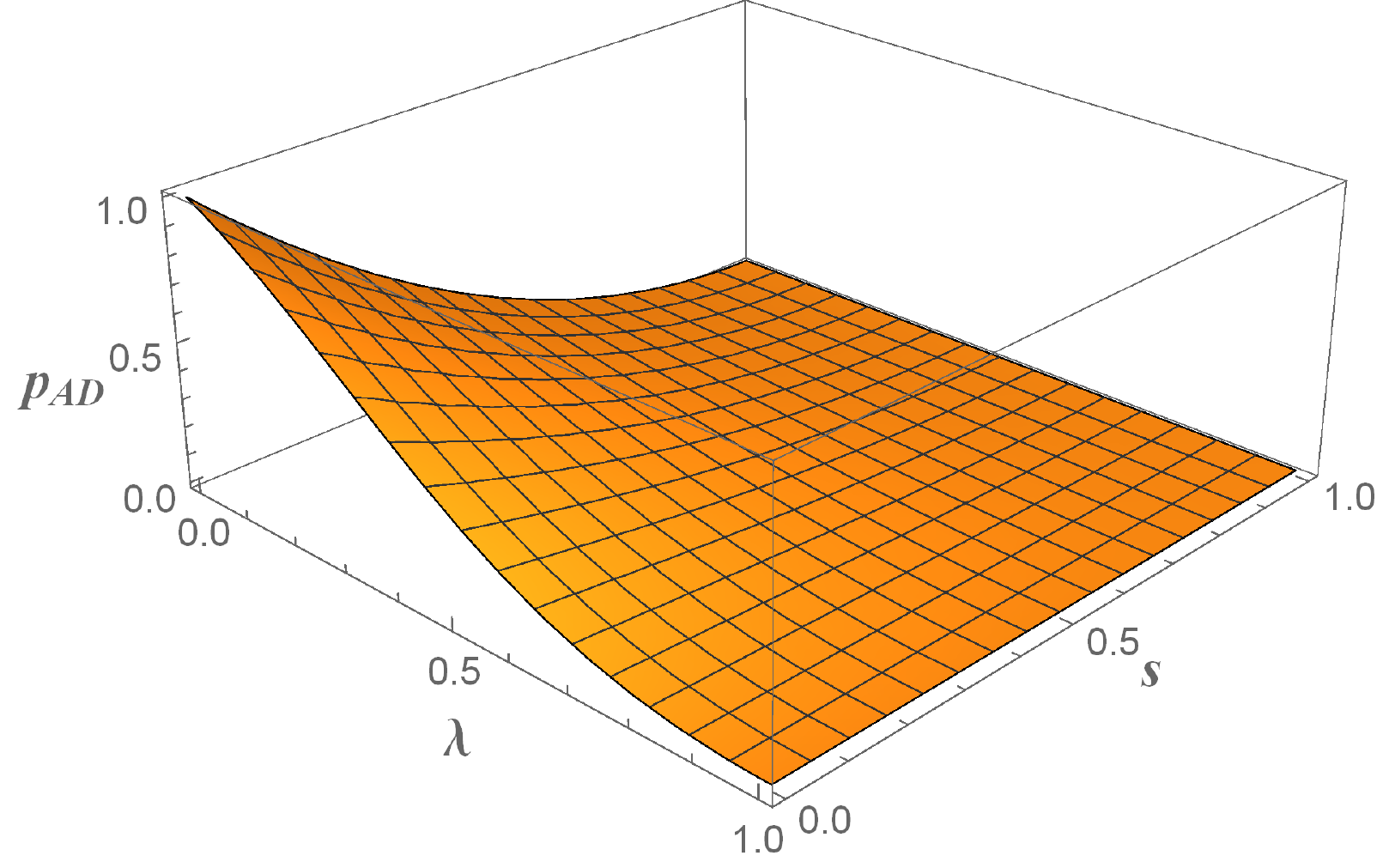}
  \caption{The success probability of the DJRSP scheme in the amplitude-damping noise with weak measurement and its reversal on condition that the strength $r$ of the reversal measurement is set to optimal.}
    \label{fig:2}
\end{figure}

The success probability in the amplitude-damping noise is plotted in Fig. \ref{fig:2}. It can be seen from the figure that the success probability $ p_{\text{AD}}$ will get 1 when there is no noise ($\lambda=0$) and no weak and reversal measurement is needed ($s = r =0$), while $ p_{\text{AD}}$ decreases with the increase of decoherence rate $\lambda$ and weak measurement strength $s$ on condition that the strength of the reversal measurement is set to $r_{\text{opt}}$.

\section{DJRSP in other types of noise with weak measurement and reversal measurement}

\subsection{In the bit-flip noise}
In the bit-flip noise, one will get the  state-independent average fidelity with weak measurement and its reversal as
\begin{equation}
\begin{split}
      \langle\overline{F'_\mathrm{BF}}\rangle = &
      \frac{1}{3 \left(r \left(\lambda ^2 r ((s-2) s+2)-2 \lambda  (r+(s-2) s)+r-2\right)+(s-2) s+2\right)}\times \\
       &~~~ \{-\lambda  \left(4 r^2+r (s (3 s-2)-6)+(s-6) s+6\right)+2 \left(r^2+r (s-3)+(s-3) s+3\right) \\
       &~~~ +\lambda ^2 (r (r ((s-2) s+3)+s (s+2)-4)-4 s+4)\}.
\end{split}
\end{equation}

Now we discuss the possibility of improving $\langle\overline{F'_\mathrm{BF}}\rangle$ in phase-flip noise by choosing the strengths of weak measurement $s$ and the reversal measurement $r$ properly.
Unfortunately,  one will get the inequation that $\langle\overline{F'_\mathrm{BF}}\rangle \leq \langle\overline{F_\mathrm{BF}}\rangle$ on condition that $0 \leq \lambda, s, r \leq 1$, where $\langle\overline{F_\mathrm{BF}}\rangle$ is in Eq. \ref{EQ22}. It is clearly that the weak measurement and its reversal cannot improve the fidelity of DJRSP in the bit-flip noise.

\subsection{In the phase-flip noise}
In the phase-flip noise, one will get the state-independent average fidelity with weak measurement and its reversal as
\begin{equation}
\begin{split}
    \langle\overline{F'_\mathrm{PF}}\rangle
      = &\frac{2 \left(r^2+4 \lambda ^2 (r-1) (s-1)-4 \lambda  (r-1) (s-1)+r s-3 r+s^2-3 s+3\right)}{3 ((r-2) r+(s-2) s+2)}.
\end{split}
\end{equation}

After careful analysis, it is shown that the fidelity $\langle\overline{F'_\mathrm{PF}}\rangle$ with weak measurements is always less than the fidelity $\langle\overline{F_\mathrm{PF}}\rangle$ in Eq. \ref{EQ23} without weak measurement. One can get the optimal $\langle\overline{F'_\mathrm{PF}}\rangle$ as
\begin{equation}
\begin{split}
 \langle\overline{F'_\mathrm{PF}}\rangle_{\text{opt}}  = \langle\overline{F_\mathrm{PF}}\rangle,
   \end{split}
\end{equation}
in the case of $r = s$. In other word, one will get $\langle\overline{F'_\mathrm{PF}}\rangle \leq \langle\overline{F_\mathrm{PF}}\rangle$ on condition that  $0 \leq \lambda, s, r \leq 1$.
It is clear that the usage of the weak measurement and its reversal cannot improve the fidelity of DJRSP in the phase-flip noise.

\subsection{In the depolarizing noise}
In the depolarizing noise, the state-independent average fidelity with weak measurement and its reversal is
\begin{equation}
\begin{split}
\langle\overline{F'_\mathrm{DE}}\rangle =
 & \frac{1}{12 \lambda ^2 r^2 ((s-2) s+2)-36 \lambda  r (r+(s-2) s)+27 ((r-2) r+(s-2) s+2)} \times \\
 &~~~   \{-6 \lambda  \left(4 r^2+r (s (3 s+2)-10)+(s-10) s+10\right)+18 \left(r^2+r (s-3)+(s-3) s+3\right) \\
 &~~~  +4\lambda ^2 (r (r ((s-2) s+3)+s (s+6)-8)-8 s+8)\}.
\end{split}
\end{equation}

\begin{figure}
\centering
\includegraphics [scale=0.55]{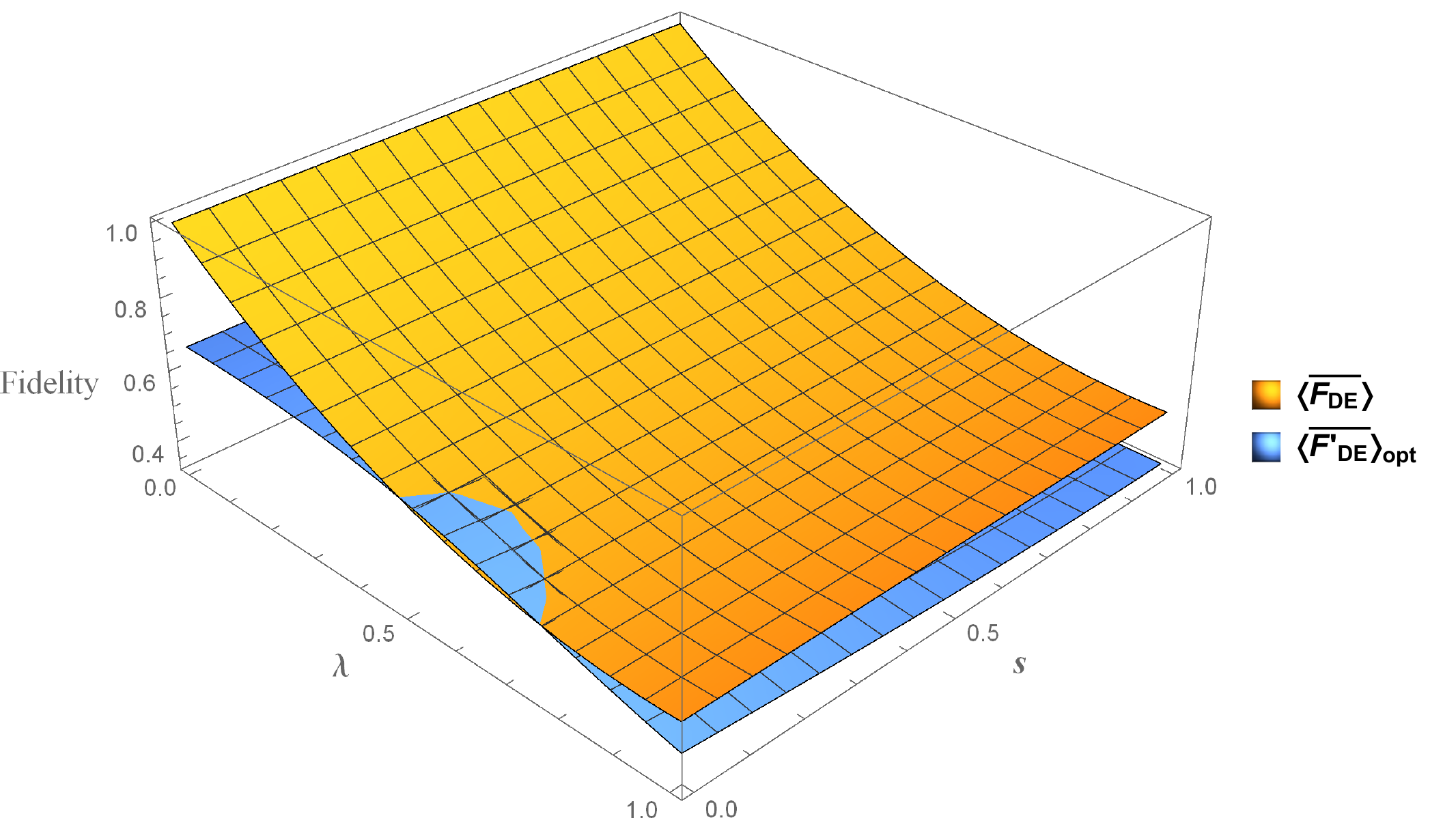}
  \caption{The state-independent average fidelities $\langle\overline{F_\mathrm{DE}}\rangle$ and $\langle\overline{F'_\mathrm{DE}}\rangle_{\text{opt}}$ with noise parameter $\lambda$ and weak measurement strength $s$ in the depolarizing noise, where $\langle\overline{F_\mathrm{DE}}\rangle$ is the fidelity without weak measurement and $\langle\overline{F'_\mathrm{DE}}\rangle_{\text{opt}}$ is the optimal  fidelity with weak measurement and its reversal. As is shown, $\langle\overline{F'_\mathrm{DE}}\rangle_{\text{opt}}$ is larger than $ \langle\overline{F_\mathrm{DE}}\rangle$ in the case of $0.468 < \lambda < 0.75$ and $0<s<1-2 \sqrt{2} \sqrt{-\frac{8 \lambda ^4-15 \lambda ^3+9 \lambda ^2}{(2 \lambda -3)^3 (8 \lambda -3)}}$.}
    \label{fig:3}
\end{figure}

The optimal value of $\langle\overline{F'_\mathrm{DE}}\rangle$ can be got as
\begin{equation}
\begin{split}
\langle\overline{F'_\mathrm{DE}}\rangle_{\text{opt}} =
\frac{2 \left(6 \lambda ^2-12 \lambda +4 \lambda ^2 s^2-12 \lambda  s^2+9 s^2-8 \lambda ^2 s+24 \lambda  s-18 s+9\right)}{3 \left(8 \lambda ^2-12 \lambda +4 \lambda ^2 s^2-12 \lambda  s^2+9 s^2-8 \lambda ^2 s+24 \lambda  s-18 s+9\right)},
\end{split}
\end{equation}
in the case of $0.468 < \lambda < 0.75$ and $0<s<1-2 \sqrt{2} \sqrt{-\frac{8 \lambda ^4-15 \lambda ^3+9 \lambda ^2}{(2 \lambda -3)^3 (8 \lambda -3)}}$,
where one can get the improvement since $\langle\overline{F'_\mathrm{DE}}\rangle > \langle\overline{F_\mathrm{DE}}\rangle$, i.e., the fidelity with weak measurement and its reversal is larger than the fidelity without measurements in Eq. \ref{EQ24}. But the maximal improvement of fidelity is less than 0.018.
$\langle\overline{F_\mathrm{DE}}\rangle$  and $\langle\overline{F'_\mathrm{DE}}\rangle_{\text{opt}}$ are plotted in Fig. \ref{fig:3}.

While in most of the others cases, $\langle\overline{F'_\mathrm{DE}}\rangle$ is less than $\langle\overline{F_\mathrm{DE}}\rangle$, which means the weak measurement and its reversal has little impact on suppressing the depolarizing noise.

\section{Conclusions}
In summary, we present a method for improving the fidelity the DJRSP through noisy channels using the technique of weak measurement and its reversal.  Taking a GHZ based DJRSP scheme as an example, we show that the loss of information about the prepared state due to interaction with noisy environment can be reduced by using weak measurement and its reversal measurement. In the amplitude-damping channel, we present the optimal strength of reversal the measurement for which the loss is minimum. Although the state-independent average fidelity can be highly improved, there is a trade-off between the improvement of fidelity and the total success probability. Surprisingly, in the bit-flip and phase-flip (phase-damping) noise, our results show that the technique of weak measurement and its reversal cannot improve the fidelity of the scheme. While in the depolarizing noise, the weak measurement only has little effect for suppressing the noise in some specific cases.

Although we have discussed the DJRSP scheme, similar results can be obtained for JRSP and RSP. To be specific, one can get simpler results for JRSP if he/she only consider the case that Alice's measurement result is $m =0$. Furthermore, one can get related results for RSP if he/she let Alice and Bob be the same person who performs two projective measurements. However, in this case, the noise effect on entanglement becomes a more simpler form since only qubit C is interacted with the noise channel. Our results will be helpful for analyzing and improving the efficiency of multiparticle entanglement based quantum communication protocols suffering from quantum noise.

\section*{Acknowledgements}
This project was supported by NSFC (Grant Nos. 61601358, 61373131), the Natural Science Basic Research Plan in Shaanxi Province of China (Program No. 2014JQ2-6030), the Scientific Research Program Funded by Shaanxi Provincial Education Department (Program No. 15JK1316), PAPD and CICAEET.


\begin{thebibliography}{10}

\bibitem{BB84}
C.H. Bennett, G.~Brassard, in \emph{Proceedings of IEEE International
  Conference on Computers Systems and Signal Processing} (IEEE, New York,
  Bangalore, India, 1984), pp. 175--179

\bibitem{HilleryBuzek-99}
M.~Hillery, V.~Bu\v{z}ek, A.~Berthiaume, Phys. Rev. A \textbf{59}(3), 1829
  (1999)

\bibitem{TerhalDivincenzo-998}
B.M. Terhal, D.P. DiVincenzo, D.W. Leung, Phys. Rev. Lett. \textbf{86}(25),
  5807 (2001)

\bibitem{CurtySantos-422}
M.~Curty, D.J. Santos, Phys. Rev. A \textbf{64}(6), 062309 (2001)

\bibitem{Wang-503}
M.M. Wang, X.B. Chen, Y.X. Yang, Sci. China Phys. Mech. Astron. \textbf{56}(9),
  1636 (2013)

\bibitem{Shamir79}
A.~Shamir, Commun. ACM \textbf{22}(11), 612 (1979)

\bibitem{XiaWang-995}
Z.~Xia, X.~Wang, X.~Sun, B.~Wang, Security and Communication Networks
  \textbf{7}(8), 1283 (2014)

\bibitem{XiaWang-996}
Z.~Xia, X.~Wang, X.~Sun, Q.~Liu, N.~Xiong, Multimed. Tools Appl.
  \textbf{75}(4), 1947 (2016)

\bibitem{MaZhou-994}
T.~Ma, J.~Zhou, M.~Tang, Y.~Tian, A.~Al-Dhelaan, M.~Al-Rodhaan, S.~Lee, IEICE
  T. Inf. Syst. \textbf{E98-D}(4), 902 (2015)

\bibitem{GuoWang-992}
P.~Guo, J.~Wang, B.~Li, S.~Lee, J. Internet Technol. \textbf{15}(6), 929 (2014)

\bibitem{RenShen-993}
Y.~Ren, J.~Shen, J.~Wang, J.~Han, S.~Lee, J. Internet Technol. \textbf{16}(2),
  317 (2015)

\bibitem{Shor97}
P.W. Shor, SIAM J. Sci. Statist. Comput. pp. 1484--1509 (1997)

\bibitem{Grover-225}
L.K. Grover, Phys. Rev. Lett. \textbf{79}(2), 325 (1997)

\bibitem{XiaWang-989}
Z.~Xia, X.~Wang, X.~Sun, Q.~Wang, IEEE T. Parall. Distr. \textbf{27}(2), 340
  (2016)

\bibitem{FuSun-1050}
Z.~Fu, X.~Sun, Q.~Liu, L.~Zhou, J.~Shu, IEICE T. Commun. \textbf{E98.B}(1), 190
  (2015)

\bibitem{FuRen-1052}
Z.~Fu, K.~Ren, J.~Shu, X.~Sun, F.~Huang, IEEE T. Parall. Distr. \textbf{27}(9),
  2546 (2016)

\bibitem{BBC-93}
C.H. Bennett, G.~Brassard, C.~Crepeau, R.~Jozsa, A.~Peres, W.K. Wootters, Phys.
  Rev. Lett. \textbf{70}(13), 1895 (1993)

\bibitem{Lo-54}
H.K. Lo, Phys. Rev. A \textbf{62}(1), 012313 (2000)

\bibitem{Pati-50}
A.K. Pati, Phys. Rev. A \textbf{63}(1), 14302 (2000)

\bibitem{BennettDivincenzo-53}
C.H. Bennett, D.P. DiVincenzo, P.W. Shor, J.A. Smolin, B.M. Terhal, W.K.
  Wootters, Phys. Rev. Lett. \textbf{87}(7), 077902 (2001)

\bibitem{Zeng-65}
B.~Zeng, P.~Zhang, Phys. Rev. A \textbf{65}(2), 022316 (2002)

\bibitem{Leung-90}
D.W. Leung, P.W. Shor, Phys. Rev. Lett.s \textbf{90}(12), 127905 (2003)

\bibitem{YeZhang-69}
M.Y. Ye, Y.S. Zhang, G.C. Guo, Phys. Rev. A \textbf{69}(2), 022310 (2004)

\bibitem{ZhouCheng-1071}
N.R. Zhou, H.L. Cheng, X.Y. Tao, L.H. Gong, Quantum Inf. Process.
  \textbf{13}(2), 513 (2014)

\bibitem{Xia-40}
Y.~Xia, J.~Song, H.S. Song, J. Phys. B: At. Mol. Opt. Phys. \textbf{40}(18),
  3719 (2007)

\bibitem{Nguyen-41}
B.A. Nguyen, J.~Kim, J. Phys. B: At. Mol. Opt. Phys. \textbf{41}(9), 095501
  (2008)

\bibitem{Hou-48}
K.~Hou, J.~Wang, Y.L. Lu, S.H. Shi, Int. J. Theor. Phys. \textbf{48}(7), 2005
  (2009)

\bibitem{Luo-283}
M.X. Luo, X.B. Chen, S.Y. Ma, X.X. Niu, Y.X. Yang, Opt. Commun.
  \textbf{283}(23), 4796 (2010)

\bibitem{Nguyen-283}
B.A. Nguyen, Opt. Commun. \textbf{283}(20), 4113 (2010)

\bibitem{XiaoLiu-386}
X.Q. Xiao, J.M. Liu, G.H. Zeng, J. Phys. B: At. Mol. Opt. Phys. \textbf{44},
  075501 (2011)

\bibitem{NguyenCao-385}
B.A. Nguyen, T.B. Cao, V.D. Nung, Phys. Lett. A \textbf{375}(41), 3570 (2011)

\bibitem{ChenXia-384}
Q.Q. Chen, Y.~Xia, J.~Song, J. Phys. A: Math. Theor. \textbf{45}, 055303 (2012)

\bibitem{Wang-471}
M.M. Wang, X.B. Chen, Y.X. Yang, Commun. Theor. Phys. \textbf{59}(5), 568
  (2013)

\bibitem{WangWang-1001}
M.M. Wang, W.~Wang, J.G. Chen, A.~Farouk, Quantum Inf. Process. \textbf{14}(1),
  4211 (2015)

\bibitem{XiangLi-941}
G.Y. Xiang, J.~Li, B.~Yu, G.C. Guo, Phys. Rev. A \textbf{72}(1), 012315 (2005)

\bibitem{Ai-XiLi-942}
A.X. Chen, L.~Deng, J.H. Li, Z.M. Zhan, Commun. Theor. Phys. \textbf{46}(2),
  221 (2006)

\bibitem{GuanChen-939}
X.W. Guan, X.B. Chen, L.C. Wang, Y.X. Yang, Int. J. Theor. Phys.
  \textbf{53}(7), 2236 (2014)

\bibitem{LiangLiu-983}
H.Q. Liang, J.M. Liu, S.S. Feng, J.G. Chen, J. Phys. B, At. Mol. Opt. Phys.
  \textbf{44}(11), 115506 (2011)

\bibitem{LiangLiu-1054}
H.Q. Liang, J.M. Liu, S.S. Feng, J.G. Chen, X.Y. Xu, Quantum Inf. Process.
  \textbf{14}(10), 3857 (2015)

\bibitem{SharmaShukla-1057}
V.~Sharma, C.~Shukla, S.~Banerjee, A.~Pathak, Quantum Inf. Process.
  \textbf{14}(9), 3441 (2015)

\bibitem{LiLiu-1056}
J.F. Li, J.M. Liu, X.Y. Xu, Quantum Inf. Process. \textbf{14}(9), 3465 (2015)

\bibitem{AlbertVaidman-1084}
Y.~Aharonov, D.Z. Albert, L.~Vaidman, Phys. Rev. Lett. \textbf{60}(14), 1351
  (1988)

\bibitem{UedaKoashi-1078}
M.~Koashi, M.~Ueda, Phys. Rev. Lett. \textbf{82}(12), 2598 (1999)

\bibitem{LeeJeong-1080}
J.C. Lee, Y.C. Jeong, Y.S. Kim, Y.H. Kim, Opt. Express \textbf{19}(17), 16309
  (2011)

\bibitem{KimLee-1075}
Y.S. Kim, J.C. Lee, O.~Kwon, Y.H. Kim, Nat. Phys. \textbf{8}(2), 117 (2012)

\bibitem{XiaAn-1076}
Z.X. Man, Y.J. Xia, N.B. An, Phys. Rev. A \textbf{86}(5), 052322 (2012)

\bibitem{PramanikMajumdar-1082}
T.~Pramanik, A.S. Majumdar, Phys. Lett. A \textbf{377}(44), 3209 (2013)

\bibitem{QiuTang-1083}
L.~Qiu, G.~Tang, X.~Yang, A.~Wang, Ann. Phys.-New. York. \textbf{350}, 137
  (2014)

\bibitem{RayChatterjee-1081}
M.~Ray, S.~Chatterjee, I.~Chakrabarty, Eur. Phys. J. D \textbf{70}(5), 114
  (2016)

\bibitem{Xian-Ting-940}
X.T. Liang, Commun. Theor. Phys. \textbf{39}(5), 537 (2003)

\bibitem{RigolinFortes-1033}
G.~Rigolin, R.~Fortes, Phys. Rev. A \textbf{92}(1), 012338 (2015)

\end{thebibliography}
\end{document}